\documentclass[%
 reprint,
 superscriptaddress,
 amsmath,amssymb,
 aps,
prb,
]{revtex4-2}

\usepackage{dcolumn}
\usepackage{bm}
\usepackage{amsmath}
\usepackage{txfonts}
\usepackage[T1]{fontenc}
\usepackage{xspace}
\usepackage{comment}
\usepackage{bbold}
\usepackage{braket}
\usepackage{ascmac}

\ifx\pdfoutput\undefined
\usepackage[dvipdfmx]{graphicx}
\usepackage[dvipdfmx]{hyperref}
\usepackage[dvipdfmx]{color}
\usepackage[dvipdfmx]{xcolor}
\else
\usepackage{graphicx}
\usepackage{hyperref}
\usepackage{color}
\usepackage{xcolor}
\fi

\newcommand{\hstv}{{\bf h}^{\rm stat}}
\newcommand{\hrotv}{{\bf h}^{\rm rot}}

\newcommand{\hst}{h^{\rm stat}}

\newcommand{\hrot}{h^{\rm rot}}
\newcommand{\hatv}[1]{\hat{\bf #1}}
\newcommand{\DNsol}{\Delta N_{\rm soliton}}

\graphicspath{
{./}
}

\begin{document}

\title{
Soliton penetration from edges in a 
monoaxial chiral magnet
}

\author{Kotaro Shimizu}
\affiliation{RIKEN Center for Emergent Matter Science, Saitama 351-0198, Japan}

\author{Shun Okumura} 
\affiliation{Department of Applied Physics, The University of Tokyo, Tokyo 113-8656, Japan}

\author{Yasuyuki Kato}
\affiliation{Department of Applied Physics, University of Fukui, Fukui 910-8507, Japan}

\author{Yukitoshi Motome}
\affiliation{Department of Applied Physics, The University of Tokyo, Tokyo 113-8656, Japan}

\date{\today}

\begin{abstract}

The solitonic spin textures such as chiral solitons, magnetic skyrmions, and magnetic hopfions, exhibiting particlelike nature widely emerge in magnets depending on spatial dimension.  
In the presence of magnetic solitons, their number directly gives rise to an impact on magnetic properties and electronic properties such as magnetoresistance and hence, it is of great importance to control the number of solitons.   
Meanwhile, a systematic study on dynamical processes to control the number of solitons, particularly by adding the desired number of solitons to the ground state exhibiting periodic arrangements of solitons, has been limited thus far. 
In this paper, we theoretically perform the systematic analysis for the dynamical control of the number of chiral solitons in 
monoaxial chiral magnets by effectively utilizing the edge modes, whose excitation is localized near the edges. 
First, by using the linear spin-wave theory, we clarify that the edge modes are brought about within the bulk magnon band gap opened by the solitonic feature introduced by the static magnetic field perpendicular to the helical axis. 
Next, by using the Landau-Lifshitz-Gilbert equation, we find the dynamical process associated with this edge mode, which exhibits the soliton penetration into the periodic arrays of chiral solitons, by applying the rotating magnetic field. 
We clarify that an additional chiral soliton penetrates the system at the right (left) edge with the precursor wherein bulk solitons move to the left (right) with the counterclockwise (clockwise) rotating magnetic field. 
Moreover, we clarify that multiple soliton penetrations can take place till the system reaches the nonequilibrium steady state, and the number of infiltrated solitons successively increases with the amplitude of the rotating magnetic field after surpassing the threshold. 
We also clarify that the static magnetic field parallel to the helical axis brings about the difference in the number of penetrating solitons as well as the threshold amplitude between the clockwise and counterclockwise rotating magnetic field. 
Our results reveal that the desired number of solitons can be added within a certain range by taking advantage of the edge modes that appear without any special processing at the edges of the system. These results contribute to the development of an experimental way to control the number of solitons and are expected to be further applied to a wide range of 
magnetic solitons, not limited to 
chiral solitons.

\end{abstract}



\maketitle



\section{Introduction \label{sec:1}}

Magnetic solitons are spin structures with particle-like nature that play a central role in magnetic and electronic properties in a wide range of magnets. 
They are often accompanied by swirling spin textures along with nontrivial topological properties, giving rise to robustness against external perturbations. 
There are a variety of magnetic solitons depending on their spatial modulation, i.e., one-dimensional chiral solitons~\cite{Dzyaloshinskii1964,Izyumov1984,Kishine2005}, two-dimensional magnetic skyrmions~\cite{Bogdanov1989,Bogdanov1994}, and three-dimensional magnetic hopfions~\cite{Sutcliffe2018,Voinescu2020,Kent2021,Rybakov2022}. 
While these magnetic solitons may appear as isolated objects, they usually form spin textures with their periodic arrangements, dubbed topological spin crystals. 
For instance, in chiral magnets with breaking of spatial inversion symmetry, the interplay between the Dzyaloshinskii-Moriya (DM) interaction and the external magnetic field perpendicular to the helical axis stabilizes a crystallization of magnetic solitons in general. 
An illustrative example is the two-dimensional skyrmion crystal, exhibiting a triangular arrangement of magnetic skyrmions in the so-called B20 compounds~\cite{Roessler2006, Muhlbauer2009}. 
In monoaxial chiral magnets such as CrNb$_3$S$_6$~\cite{Togawa2012} and Yb(Ni$_{1-x}$Cu$_{x}$)$_3$Al$_9$~\cite{Matsumura2017}, a one-dimensional periodic arrangement of chiral solitons, called the chiral soliton lattice (CSL), appears in a magnetic field~\cite{Dzyaloshinskii1964,Izyumov1984,Kishine2005,Togawa2012}, which is a main focus of this study.

In the topological spin crystals, creation and annihilation of magnetic solitons in the bulk require considerable energy cost, thereby underscoring the importance of edges and surfaces of the system to control the number of solitons. 
Particularly, in the 
monoaxial chiral magnets, it was shown that chiral solitons experience a potential barrier at the edges, which depends on the external magnetic field~\cite{Shinozaki2018}. 
Consequently, the process of adding and removing a soliton at the edges exhibits nonreciprocity, causing a large hysteresis in the field dependence of the magnetization as well as the magnetoresistance, with discrete jumps corresponding to the changes of soliton number~\cite{Wilson2013,Kishine2014,Togawa2015,Wang2017, Shinozaki2018,Ohkuma2019,Paterson2019}. 
Thus, clarifying nonequilibrium processes near the edges is crucially important for understanding of magnetic and electronic properties of the system.

The central question we address in this study is how one can generate chiral solitons at the edges efficiently, especially in the CSL state, namely, in the existence of a periodic array of solitons. 
Thus far, the soliton generation has been investigated mainly for the process of creating a single or a few solitons in the forced ferromagnetic (FFM) state, corresponding to the vacuum of solitons, near the saturation field. 
Such a process can be achieved by, e.g., modulating the potential barrier at the edges with the external magnetic field~\cite{Du2015,Muller2016} and rotating the spins near the edges~\cite{Schaffer2020}. 
Nevertheless, generation processes in the CSL state are more intriguing since they affect the magnetic and electronic properties over a wider range of the magnetic field.  
Particularly, it would be of great interest to clarify the dynamical processes under time-dependent magnetic fields for establishing experimentally feasible control of the soliton number.

In this paper, to address these challenges, we theoretically study the dynamical control of soliton number in a 
monoaxial chiral magnet through the edge modes, which are magnetic excitation modes localized near the edges~\cite{Masaki2018, Masaki2018Instabilities, Hoshi2020, Shimizu2023EEF}. 
First, by using the linear spin-wave theory, we calculate the magnon spectrum exhibiting the band gap opened by the solitonic feature of the spin texture in an external magnetic field perpendicular to the helical axis. 
We show that, in the presence of edges, additional edge modes appear within the magnon band gaps. 
We also show that the frequency of the edge mode increases with the magnetic field up to the saturation to the FFM state. 
Next, by numerical simulations based on the Landau-Lifshitz-Gilbert (LLG) equation, we find intriguing dynamics exhibiting the soliton penetration from the edges of the system by applying the rotating magnetic field with the frequency of the edge mode. 
We clarify that the spins at the left and right edges exhibit precession motion, yet simultaneously, bulk solitons start moving to the left or right depending on the rotating direction. 
We reveal that after the system exhibits these precursor motions, additional solitons penetrate into the system till the system reaches the nonequilibrium steady state. 
The number of infiltrated solitons successively increases with the amplitude of the rotating magnetic field surpassing the threshold value. 
Furthermore, we also show that the thresholds for the opposite direction of rotation become inequivalent when the static magnetic field breaks the symmetry by its parallel component to the helical axis.

The rest of the paper is organized as follows. 
In Sec.~\ref{sec:2}, we introduce the model (Sec.~\ref{sec:model}), the numerical methods of the linear spin-wave theory (Sec.~\ref{sec:LSWT}) and the LLG equation (Sec.~\ref{sec:LLGeq}). 
In Sec.~\ref{sec:SW}, we present the results obtained with the linear spin-wave theory. 
We first show the magnon band structure for the CSL state in the bulk (Sec.~\ref{sec:SW_band}), and then, present the spin-wave spectrum and real-space distribution of the magnon wave function calculated in the presence of edges (Sec.~\ref{sec:SW_edge}). 
In Sec.~\ref{sec:penetration}, we present the results of the real-time evolution obtained by solving the LLG equation, associated with the irradiation of the rotating magnetic field with the frequency of the edge mode. 
We show the spatiotemporal profiles of the spin texture and the number of penetrating solitons by changing the amplitude of the rotating magnetic field in the presence of the static magnetic field perpendicular to the helical axis (Sec.~\ref{sec:penetration_hx=0}) followed by the results with the parallel field component (Sec.~\ref{sec:penetration_hx>0}). 
We discuss the results in Sec.~\ref{sec:discussion}. Section~\ref{sec:summary} is devoted to the summary of this paper.



\section{Model and method \label{sec:2}}

In this section, we introduce the model for a monoaxial chiral magnet and the numerical methods to investigate the spin-wave excitation and the dynamics of the model. 
In Sec.~\ref{sec:model}, we introduce the model Hamiltonian and the setting of parameters used in this study. 
We describe the numerical methods based on the linear spin-wave theory and the LLG equation in Secs.~\ref{sec:LSWT} and \ref{sec:LLGeq}, respectively.

\subsection{Model \label{sec:model}}

\begin{figure}[tb]
\centering
\includegraphics[width=1.0\columnwidth]{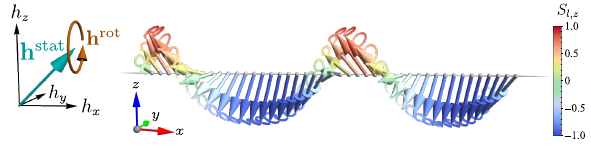}
\caption{
\label{fig:setup}
Schematic picture of the setup in this study. 
The arrows on the chain in the $x$ direction represent the spin configuration in the model in Eq.~\eqref{eq:ham}; the color denotes the $S_z$ component of spins. 
The pale curves attached to the arrowheads represent the trajectories of the spins in the time evolution under the static magnetic field $\hstv$ and the rotating magnetic field $\hrotv$ depicted in the inset. 
}
\end{figure}

In canonical examples of monoaxial chiral magnets hosting the CSL, such as CrNb$_3$S$_6$, spin interactions within the two-dimensional planes perpendicular to the helical axis are strong, while those between different planes are approximately an order of magnitude weaker~\cite{Togawa2016,Shinozaki2016}. 
Particularly at zero temperature, spins in each plane can be considered fully aligned, and the spin dynamics can be effectively described by the time-dependent one-dimensional Hamiltonian~\cite{Togawa2016} 
\begin{eqnarray}
\mathcal{H}(t)=
\sum_l && \bigl[ -J{\bf S}_l(t)\cdot{\bf S}_{l+1}(t) 
-D\hat{\bf x} \cdot \left( {\bf S}_l(t) \times {\bf S}_{l+1}(t) \right) \notag \\ 
&& ~+ {\bf h}(t)\cdot{\bf S}_l(t) \bigr], 
\label{eq:ham}
\end{eqnarray}
where ${\bf S}_l(t)$ represents the classical spin at site $l$ along the helical axis and time $t$ with $|{\bf S}_l(t)|=1$. 
The first and second terms in the square brackets denote the Heisenberg and Dzyaloshinskii-Moriya (DM) interactions, respectively. 
The last term describes the Zeeman coupling to a time-dependent external magnetic field ${\bf h}(t)$; we use the convention of a positive sign. 
The magnetic field consists of the static and rotating parts, denoted by ${\bf h}^{\rm stat}$ and ${\bf h}^{\rm rot}(t)$, respectively, as 
\begin{eqnarray}
{\bf h}(t)={\bf h}^{\rm stat}+{\bf h}^{\rm rot}(t). 
\label{eq:def_magneticfield}
\end{eqnarray}
See Fig.~\ref{fig:setup} for the setup of the model. 
In this study, without loss of generality, the static magnetic field is applied in the $xz$ plane, i.e., $\hstv=(\hst_x, 0, \hst_z)$, since the model has continuous rotational symmetry about the $x$ axis. 
The explicit form of ${\bf h}^{\rm rot}(t)$ will be given later [Eq.~\eqref{eq:hrot}]. 

\begin{figure}[tb]
\centering
\includegraphics[width=1.0\columnwidth]{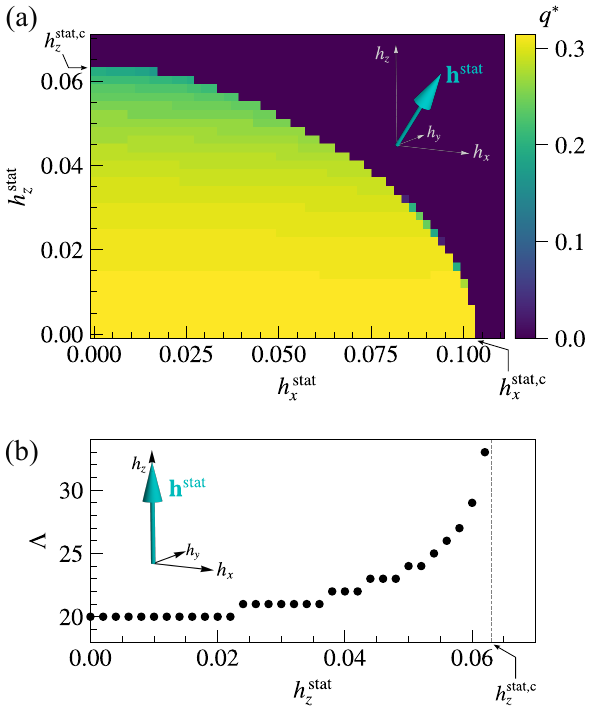}
\caption{
\label{fig:phase_diagram}
(a) Phase diagram for the Hamiltonian in Eq.~\eqref{eq:ham} with $\hrotv(t)=0$ on the plane of $\hst_x$ and $\hst_z$. 
The color represents the ordering wave number $q^*$ calculated for the $L=1000$ system; 
$q^*=\frac{\pi}{10}$ for the helical state at $\hst_x=\hst_z=0$ and the conical states for $0<\hst_x<h_x^{\rm stat,c}\simeq 0.103$ at $\hst_z=0$, $q^*=0$ for the FFM state above the critical field curve connecting $h_x^{\rm stat,c}$ and $h_z^{\rm stat,c}\simeq 0.063$, and $0<q^*\leq\frac{\pi}{10}$ otherwise. 
(b) The magnetic modulation period for the commensurate approximant of the stable spin configuration, $\Lambda$, for $\hst_x=0$. 
}
\end{figure}

Hereafter, we set the energy scale as $J=1$ and the lattice constant as unity, and take $D=\tan\left(\frac{\pi}{10}\right)$. 
In a static case 
for ${\bf h}^{\rm rot}(t)=0$
, the model in Eq.~\eqref{eq:ham} is time independent and exhibits an incommensurate magnetic state below the critical field above which the FFM state is stable. 
The spin structure depends on the amplitude and the direction of ${\bf h}^{\rm stat}$. 
The ordering wave number for the ground state, $q^*$, is shown in Fig.~\ref{fig:phase_diagram}(a) on the plane of $\hst_x$ and $\hst_z$. 
The result is obtained by minimizing the energy in the system with $L=1000$ sites under the periodic boundary condition (PBC). 
In this paper, the energy minimization is done by using the time evolution by the LLG equation in Sec.~\ref{sec:LLGeq} or employing Adam~\cite{Kingma2014adam} by using the libraries Optax~\cite{deepmind2020jax} and JAX~\cite{jax2018github}, both of which give essentially the same results. 
In the absence of the magnetic field, the ground state is a helical spin structure, in which the spins rotate in the $yz$ plane with a period of $20$ lattice sites, i.e., $q^*=\arctan\left(\frac{D}{J}\right)=\frac{\pi}{10}$. 
When the static magnetic field is introduced, the model stabilizes a conical spin structure, a CSL, and their mixture, in addition to the FFM state with $q^*=0$ above the critical field, depending on the direction and the strength of the field~\cite{Dzyaloshinskii1964, Izyumov1984, Kishine2005, Victor2016, Masaki2018}. 
By increasing $\hst_x$ with $\hst_z=0$, the spins on the $yz$ plane in the zero-field helical state exhibit a uniform spin canting in the $x$ direction to form the conical spin structure without changing $q^*$ up to the FFM transition at $h_{x}^{\rm stat, c}=2\left(\sqrt{J^2+D^2}-J\right)\simeq0.103$. 
In contrast, by increasing $\hst_z$ with $\hst_x=0$, a solitonic feature is introduced in the spin structure and the magnetic modulation period increases ($q^*$ decreases) to form the CSL up to the FFM transition at the critical field $h_z^{\rm stat, c}\simeq0.063$. 
We note that the critical field for our lattice model in Eq.~\eqref{eq:ham} is slightly different from the analytical value in the continuum limit $\left(\frac{\pi}{4} \arctan\left(\frac{D}{J}\right)\right)^2J\simeq0.061$ due to the lattice discretization effect~\cite{Togawa2016}. 
When both $\hst_x$ and $\hst_z$ are nonzero, a complicated spin texture with a mixture of conical and CSL is stabilized.

In the following calculations, we take two different boundary conditions depending on the analysis. 
For the magnon band calculation in Sec.~\ref{sec:SW_band}, we take the PBC while varying the system size $L$ depending on the magnetic modulation period of the ground state spin texture (see Sec.~\ref{sec:LSWT} for the details). 
Meanwhile, to clarify the edge contribution to the spin-wave excitation in Sec.~\ref{sec:SW_edge}, we perform the linear spin-wave analyses under the open boundary condition (OBC), in addition to the PBC, for the system size of $L=1000$. 
In Sec.~\ref{sec:penetration}, we calculate the spin dynamics under the OBC with $L=1000$. 
The sum in Eq.~\eqref{eq:ham} is taken from $l=0, 1, \ldots, L-1$, 
where we impose ${\bf S}_{L}(t)={\bf S}_0(t)$ for the PBC, whereas ${\bf S}_{L}(t)=0$ for the OBC.


\subsection{Linear spin-wave theory \label{sec:LSWT}}

To clarify the spin-wave excitation in the static model given by Eq.~\eqref{eq:ham} with ${\bf h}^{\rm rot}(t)=0$, we study the magnon spectrum and the spatial profile of the excitation modes by using the linear spin-wave theory; technical details can be found, e.g., in Refs.~\cite{Colpa1978,Toth2015,Kato2021}. 
For the stable spin configuration in the ground state, we introduce a new local spin frame where 
the spin configuration is regarded as the collinear ferromagnetic state. 
The spin in the new frame is denoted by $\tilde{\bf S}_l$ and the stable spin configuration is represented by $\tilde{\bf S}_l=(0,0,1)^{{\mathsf T}}$, where ${\mathsf T}$ represents the transpose of a vector. 
By using the Holstein-Primakoff transformation, $\tilde{\bf S}_l$ is represented by the boson operators up to the leading order as
\begin{eqnarray}
\left(\begin{array}{c}
\tilde{S}_{l,x}\\
\tilde{S}_{l,y}\\
\tilde{S}_{l,z}\\
\end{array}\right)
\simeq
\left(\begin{array}{c}
\sqrt{\frac{1}{2}}(\hat{a}^{\;}_l + \hat{a}_l^{\dag})\\
-i\sqrt{\frac{1}{2}}(\hat{a}^{\;}_l - \hat{a}_l^{\dag})\\
1-\hat{a}_l^{\dag}\hat{a}^{\;}_l\\
\end{array}\right), 
\label{eq:HPtrans}
\end{eqnarray}
where $i$ is the imaginary unit 
and $\hat{a}^{\;}_l$ ($\hat{a}^{\dag}_l$) represents the annihilation (creation) operator of the magnon at site $l$.
By substituting Eq.~\eqref{eq:HPtrans} into Eq.~\eqref{eq:ham}, we obtain the magnon Hamiltonian expressed as 
\begin{eqnarray}
\mathcal{H}_{\rm SW}=\sum_{lm}\hat{\bf a}_l^{\dag}{\mathsf K}^{\;}_{lm}\hat{\bf a}^{\;}_m, 
\label{eq:HSW_real}
\end{eqnarray}
where $\hat{\bf a}^{\;}_l=\left(\hat{a}^{\;}_l, \hat{a}_l^{\dag}\right)^{\mathsf T}$. 
We note that ${\mathsf K}_{lm}$ is a $2\times2$ matrix being the $lm$ block of the $2L\times2L$ Hamiltonian matrix ${\mathsf H}$. 
The eigenenergy of the magnon, $\omega_n$ ($n=1,2,\ldots,L$), appears twice in $2L$ eigenenergies obtained by numerically diagonalizing the matrix ${\mathsf H}$~\cite{Colpa1978}.  

In general, the Hamiltonian in Eq.~\eqref{eq:ham} exhibits an incommensurate ground state. 
We calculate the corresponding magnon band structure for the spin texture by setting the magnetic modulation period of an integer $\Lambda$, which approximates the true modulation period, in the system with $L=M\Lambda$ sites under the PBC, where $M$ is also an integer. 
Here, we optimize the spin configuration to minimize the energy of the model in Eq.~\eqref{eq:ham} with ${\bf h}^{\rm rot}(t)=0$ by changing $\Lambda$ from 20 to 40. 
The spin structure with the period of $\Lambda$ which gives the lowest energy can be regarded as the commensurate approximant of the incommensurate ground-state spin texture in the thermodynamic limit. 
By increasing $\hst_z$, $\Lambda$ monotonically increases, as shown in Fig.~\ref{fig:phase_diagram}(b), which is consistent with Fig.~\ref{fig:phase_diagram}(a). 
By regarding this spin structure as a magnetic unit cell, we introduce the Fourier transformation as
\begin{eqnarray}
\hat{\bf a}^{\;}_{\alpha, q}
=\frac{1}{M}\sum_{m=0}^{M-1}\hat{{\bf a}}^{\;}_{\alpha+m\Lambda}e^{-iqm\Lambda} 
=\left(\begin{array}{c} \hat{a}^{\;}_{\alpha, q} \\ \hat{a}_{\alpha,-q}^{\dag} \end{array}\right), 
\end{eqnarray}
where the subscript $\alpha=0,1,\ldots,\Lambda-1$ denotes the coordinate in the magnetic unit cell and $q$ is the wave number. 
By Fourier transforming Eq.~\eqref{eq:HSW_real}, we obtain 
\begin{eqnarray}
\mathcal{H}_{\rm SW}=\sum_{q}\sum_{\alpha\beta}
\hat{\bf a}_{\alpha,q}^{\dag}{\mathsf K}^{\;}_{\alpha\beta,q}\hat{\bf a}^{\;}_{\beta, q}, 
\label{eq:HSW_Fourier}
\end{eqnarray}
where ${\mathsf K}_{\alpha\beta, q}$ is a $2\times2$ matrix being the $\alpha\beta$ block of the $2\Lambda\times2\Lambda$ Hamiltonian matrix ${\mathsf H}_q$. 
Finally, by diagonalizing ${\mathsf H}_q$, the $p$-th magnon band dispersion with the wave number $q$, $\omega_{p,q}$ ($p=1,2,\ldots,\Lambda$), is obtained~\cite{Colpa1978}.


\subsection{Landau-Lifshitz-Gilbert equation \label{sec:LLGeq}}

We study the real-time dynamics of the model in Eq.~(\ref{eq:ham}) by using the LLG equation given by~\cite{Landau1935,Gilbert1955} 
\begin{eqnarray}
\frac{\partial{\bf S}_l(t)}{\partial t}=\frac{1}{1+\alpha^2}&& \left[
-{\bf S}_l(t)\times{\bf h}^{\rm eff}_l(t) \right. \notag \\
&& \ \left.+\alpha {\bf S}_l(t)\times\left({\bf S}_l(t)\times{\bf h}^{\rm eff}_l(t)\right)\right], 
\label{eq:LLG}
\end{eqnarray}
where $\alpha$ is the Gilbert damping and ${\bf h}^{\rm eff}_l(t)$ is the effective magnetic field at time $t$ defined by 
\begin{eqnarray}
{\bf h}^{\rm eff}_l(t) = 
\frac{\partial \mathcal{H}(t)}{\partial {\bf S}_l(t)} 
&=& 
-J\bigl({\bf S}_{l+1}(t)+{\bf S}_{l-1}(t) \bigr) \notag \\
&&+ D\hat{\bf x}\times \bigl( {\bf S}_{l+1}(t) - {\bf S}_{l-1}(t)\bigr) + {\bf h}(t). 
\label{eq:heff}
\end{eqnarray}
Note that the length constraint of $|{\bf S}_l(t)|=1$ is deferred only when taking the derivative of the Hamiltonian with respect to ${\bf S}_l(t)$. 
Here, we impose ${\bf S}_{-1}(t)={\bf S}_{L-1}(t)$ and ${\bf S}_{-1}(t)=0$ for the PBC and OBC cases, respectively (see the last paragraph of Sec.~\ref{sec:model}). 
We numerically solve Eq.~(\ref{eq:LLG}) by using the fourth-order Runge-Kutta method with a time step $\Delta t=0.1$. 
In the following calculations, we take $\alpha=0.04$, which is a typical value for ferromagnetic metals~\cite{Mizukami2001, Mizukami2010,Oogane2006,Oogane2010}. 

In this work, the spin dynamics is driven by the rotating magnetic field with a frequency $\omega$ given by 
\begin{eqnarray}
{\bf h}^{\rm rot}(t)=h^{\rm rot}
\left(\hatv{y}\cos(\omega t) + \epsilon \hatv{z}\sin(\omega t)\right)\
\left(1-e^{-(t/t_0)^2}\right), 
\label{eq:hrot}
\end{eqnarray}
where $h^{\rm rot}$ and $\epsilon=\pm1$ represent the amplitude and the direction of rotation of the magnetic field, respectively. 
We call ${\bf h}^{\rm rot}(t)$ with $\epsilon=1$ ($-1$) the counterclockwise (clockwise) rotating magnetic field. 
In Eq.~\eqref{eq:hrot}, we introduce the exponential damping factor with $t_0=1000$ to suppress the initial disturbance by switching on the field. 
In the following results, we calculate the spin dynamics up to the time $t_f=2\times10^5$.




\section{Linear spin-wave analysis \label{sec:SW}}

In this section, we present the bulk and edge properties of the spin-wave excitation based on the linear spin-wave theory. 
In Sec.~\ref{sec:SW_band}, we show the magnon band structure for the helical and the CSL states by changing $\hst_z$ with $\hst_x=0$. 
In Sec.~\ref{sec:SW_edge}, we show the spin-wave spectrum under the PBC and OBC, and also present the real-space distribution of the magnon wave function of the edge mode.

\subsection{Magnon band structure \label{sec:SW_band}}

\begin{figure}[h]
\centering
\includegraphics[width=1.0\columnwidth]{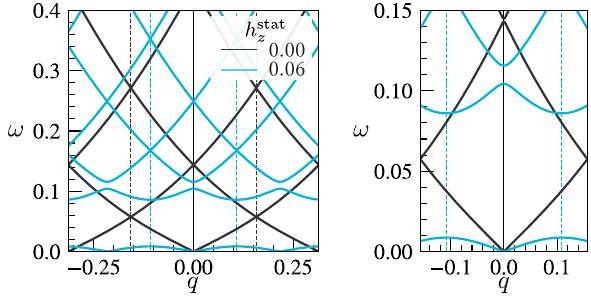}
\caption{
\label{fig:band}
The magnon band dispersion for $\hst_z=0$ and $\hst_z=0.06$ represented by the black and blue curves, respectively, at $\hst_x=0$.  
The results are shown in the repeated BZ scheme up to the second and the first magnetic BZs of $\hst_z=0$ in the left and right panels, respectively. 
The BZ edges for $\hst_z=0$ and $\hst_z=0.06$ are represented by the black and blue vertical dashed lines, respectively. 
}
\end{figure}

We show the magnon band dispersions $\omega_{p, q}$ obtained by the linear spin-wave theory in the repeated Brillouin zone (BZ) scheme in Fig.~\ref{fig:band}. 
Let us first discuss the results for the helical state with $(\hst_x,\hst_z)=(0,0)$ represented by the black lines. 
In this state, the magnetic modulation period of the ground state in the thermodynamic limit is exactly 
$\Lambda=20$, as mentioned in Sec.~\ref{sec:model}. 
The band structure in the extended BZ scheme is analytically given by
\begin{eqnarray}
\omega_{q}=2\sqrt{\tilde{J}\left(1-\cos q\right)\left(\tilde{J}-J\cos q\right)}, 
\end{eqnarray}
where $\tilde{J}=\sqrt{J^2+D^2}$~\cite{Kataoka1987, Maleyev2006,Kishine2009}. 
The dispersion is continuous without showing a gap, similar to the electron band dispersion with the empty lattice approximation. 
In addition, the helical state exhibits a Goldstone mode with a linear dispersion around $q=0$ due to the continuous rotational symmetry of the helical spin structure about the helical axis.

Next, we discuss the results for the CSL with $(\hst_x,\hst_z)=(0,0.06)$ represented by the blue lines in Fig.~\ref{fig:band}. 
In this case, the magnetic modulation period is increased up to $\Lambda=29$, as shown in Fig.~\ref{fig:phase_diagram}(b). 
We observe gap opening in the magnon dispersion; the gaps between the $(2n-1)$-th and $2n$-th bands, and between the $2n$-th and $(2n+1)$-th bands, appear at $q=\pm\frac{\pi}{\Lambda}$ and $q=0$, respectively, where $n$ represents an integer~\cite{Kishine2009}. 
This is due to the solitonic feature of the spin texture introduced by $\hst_z$, acting as a periodic potential that scatters the magnons. 
The gap magnitude is enhanced with increasing $\hst_z$, and accordingly, the width of each magnon band is reduced. 
We note that the CSL also exhibits a linear Goldstone mode around $q=0$ in the lowest branch, which corresponds to the optical phonon mode of the CSL~\cite{Izyumov1985, Bostrem2008_transport, Bostrem2008_theory,Kishine2009}.

\subsection{Edge mode \label{sec:SW_edge}}

\begin{figure}[tb]
\centering
\includegraphics[width=1.0\columnwidth]{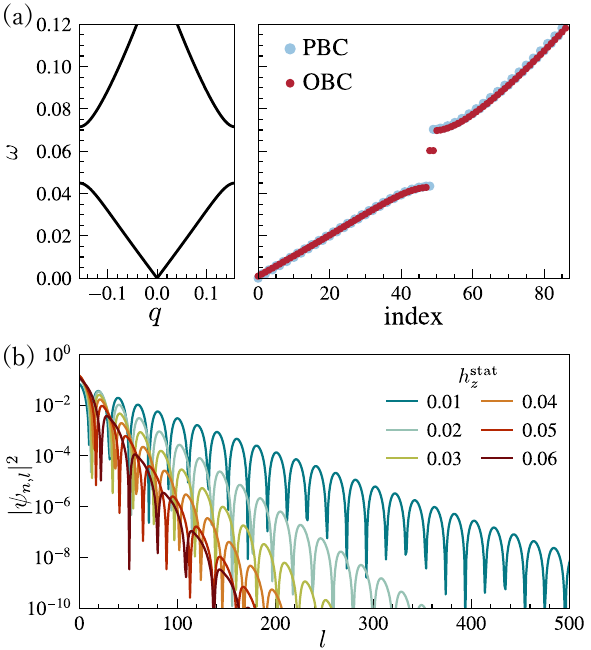}
\caption{
\label{fig:edge_spectrum}
(a) Magnon band dispersion for the first BZ (left) and eigenfrequencies of the magnon Hamiltonian for the 1000-site system under the PBC and OBC shown by the blue and red points, respectively (right). The results are shown for $(\hst_x, \hst_z)=(0, 0.02)$. 
In the right panel, the eigenfrequencies are shown in ascending order. 
(b) Real-space distribution of the square amplitude of the eigenvectors $|\psi_{n,l}|^2$ at $\hst_x=0$ for several values of $\hst_z$.
For each $\hst_z$, the eigenfrequency is set at the edge mode localized at the left edge of the system. 
}
\end{figure}

To clarify the edge effects on the spin-wave excitation, we compare the magnon spectrum between the PBC and OBC cases. 
Figure~\ref{fig:edge_spectrum}(a) shows the magnon band structure and the eigenfrequencies of the real-space magnon Hamiltonian in Eq.~\eqref{eq:HSW_real} for the $L=1000$-site system with $(\hst_x, \hst_z)=(0, 0.02)$. 
The magnon band dispersion obtained for the PBC case is shown in the left panel, which exhibits a gap for $0.045 \lesssim \omega \lesssim 0.07$ due to nonzero $\hst_z$. 
The eigenfrequencies are shown in the right panel in ascending order. 
We also plot the results for the OBC case, which almost overlap with the PBC ones, except for two additional modes appearing in the band gap. 
These mid-gap states are edge modes~\cite{Hoshi2020}.  
Indeed, 
the intensity of the corresponding eigenvectors $\psi_{n,l}$ are localized around the edges of the system, as shown in Fig.~\ref{fig:edge_spectrum}(b) for one of the two modes localized at the left edge (the other is localized at the right edge). 
As shown in Fig.~\ref{fig:edge_spectrum}(b), these edge modes are more localized for stronger magnetic field $\hst_z$. 
This is consistent with the increase of the band gap in the bulk spectrum, shown in Fig.~\ref{fig:band}. 
We note that two eigenfrequencies are degenerate in the ideal case due to the symmetry of the static Hamiltonian in Eq.~\eqref{eq:ham} with the twofold rotational operation about the $y$ axis followed by the time reversal operation; however in our numerically optimized solutions, the degeneracy is slightly lifted, allowing the two modes localized at the left and right edges to be resolved.

\begin{figure}[tb]
\centering
\includegraphics[width=1.0\columnwidth]{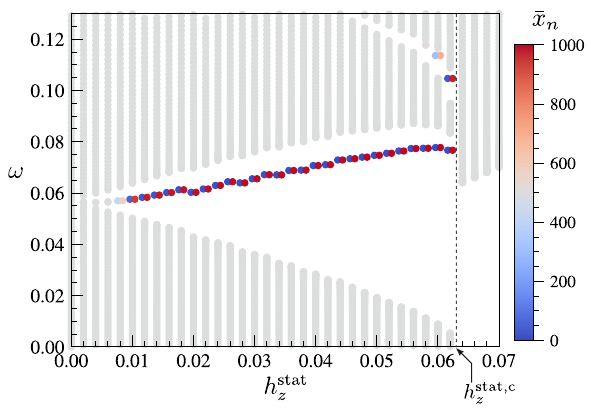}
\caption{
\label{fig:edge_hz}
Eigen frequencies of the magnon Hamiltonian with varying $\hst_z$ at $\hst_x=0$. 
The color represents the center of mass of the eigenmodes [Eq.~\eqref{eq:xbar}]. 
For a better visibility, each edge mode pair is plotted slightly offset to the left and right.
}
\end{figure}

\begin{figure*}[tb]
\centering
\includegraphics[width=2.0\columnwidth]{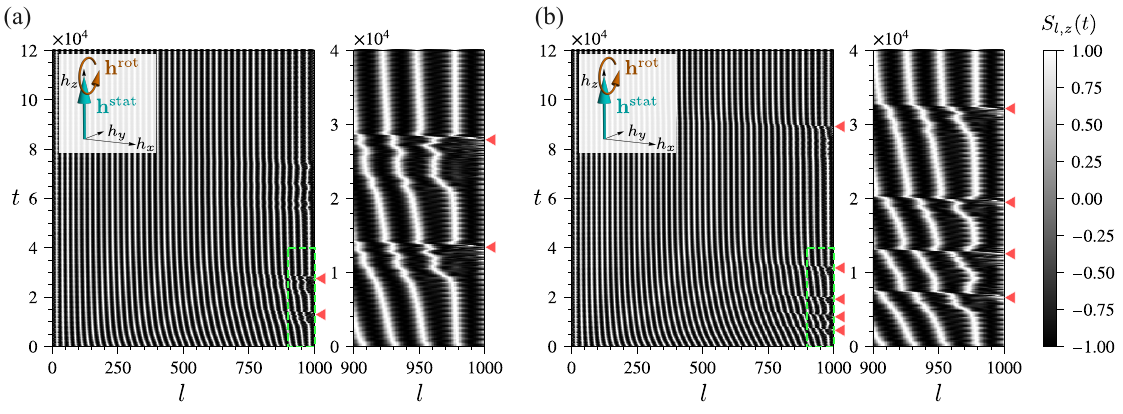}
\caption{
\label{fig:spin_xt}
Real-space and real-time distributions of the $z$ component of spins $S_{l,z}(t)$ on the plane of $l$ and $t$ for the counterclockwise rotating magnetic field ($\epsilon=1$) at $(\hst_x, \hst_z)=(0, 0.05)$ and $\omega=0.075$:  (a) $h^{\rm rot}=0.01$ and (b) $h^{\rm rot}=0.012$. 
The frequency $\omega$ is set at the edge modes in the first gap (see Fig.~\ref{fig:edge_hz}). 
The red triangles represent the time when solitons penetrate from the right edge of the system. 
In each panel, the left figure shows the whole system, while the right one is an enlarged plot near 
the right edge in a shorter time window, denoted by the dashed green square in the left one. 
}
\end{figure*}

In Fig.~\ref{fig:edge_hz}, we show the $\hst_z$ dependence of the eigenspectra at $\hst_x=0$ under the OBC. 
Each dot represents the eigenfrequency $\omega_n$ and the color shows the center of mass of the eigenmode, defined by
\begin{eqnarray}
\bar{x}_n
=\frac{\sum\limits_{l=0}^{L-1} l\left|\psi_{n,l}\right|^2}{\sum\limits_{l=0}^{L-1} \left|\psi_{n,l}\right|^2}. 
\label{eq:xbar}
\end{eqnarray}   
The gray dots represent the bulk modes spreading over the system, whereas the blue and red dots denote the edge modes localized at the left and right edges of the system, respectively. 
We find that the frequency of the edge mode appearing in the lowest-energy band gap increases with $\hst_z$, except for the very vicinity of the critical field $h_z^{\rm{stat, c}}$. 
We also observe the edge modes within the second gap, while they are obscure compared to the ones in the first gap. 
The other edge modes in the higher-energy gap are hard to identify since the gaps become smaller. 
We note that these localized modes appear at not only the edges of the system but also the boundaries of domains with different chirality (see Appendix~\ref{sec:penetration_domain}).



\section{Edge excitation and soliton penetration \label{sec:penetration}}

In the preceding section, we have elucidated the characteristics of the magnetic excitation spectrum in the static magnetic field, particularly focusing on the edge modes. 
In this section, we show the intriguing dynamics wherein the edge modes are vigorously excited, leading to a progressive increase in the number of solitons over time by successive penetration from the edges.
In Sec.~\ref{sec:penetration_hx=0}, we present the results with $\hst_x=0$, 
delving into the spatiotemporal behavior of the spins, and the relationship between the amplitude of the rotating magnetic field and the number of solitons penetrating into the system. Subsequently, in Sec.~\ref{sec:penetration_hx>0}, we extend the analysis to the case with $\hst_x\neq0$, thereby showcasing the effects introduced by the static magnetic field parallel to the helical axis.

\subsection{$h_x=0$ case \label{sec:penetration_hx=0}}

First, we show the results of the spatiotemporal profiles of the spin configurations ${\bf S}_l(t)$ by numerically solving the LLG equation given in Eq.~\eqref{eq:LLG} in the absence of the static magnetic field parallel to the helical axis, $\hst_x$. 
For the dynamical simulations, we use the ground-state spin configuration obtained by numerically solving the LLG equation with $\hrotv(t)=0$ as the initial state.
Figure~\ref{fig:spin_xt} displays the real-space and real-time distribution of $S_{l,z}(t)$ at $(\hst_x,\hst_z)=(0, 0.05)$ under the rotating magnetic field in Eq.~\eqref{eq:hrot} with $\epsilon=1$ and $\omega=0.075$. 
The frequency $\omega$ is set at the edge modes in the first gap (see Fig.~\ref{fig:edge_hz}). 
In each figure of Fig.~\ref{fig:spin_xt}, the left panel illustrates the temporal evolution up to $t=1.2\times 10^5$ across all sites of the 1000-site system, while the right panel depicts an enlarged plot of the rightmost 100 sites near the right edge of the system in a shorter time up to $t=4\times10^4$.

Figure~\ref{fig:spin_xt}(a) is for the result with $h^{\rm rot}=0.01$. 
By applying the rotating magnetic field with the frequency of the edge mode, 
the spins located at the left and right edges experience substantial modulations and exhibit precession motion. 
Simultaneously, solitons within the bulk start moving to the left, as shown in the left panel of Fig.~\ref{fig:spin_xt}(a). 
This is a consequence of the breaking of twofold rotational symmetry about the $z$ axis by the rotating magnetic field, leading to the unidirectional propagation of the spin wave.  
This motion compresses the soliton lattice toward the left edge, leaving soliton-sparse regions near the right edge where spins predominantly point downward (in the $-\hatv{z}$ direction). 
Then, the system exhibits a pronounced precursor phenomenon showing the rapid shift of the rightmost soliton at $t \simeq 0.9\times10^4$, and subsequently causes a soliton infiltration from the right edge into the system at $t \simeq 1.4 \times 10^4$, as marked by the lower red triangle; see the right panel of Fig.~\ref{fig:spin_xt}(a). 
Another soliton penetration takes place similarly at $t \simeq 2.8 \times 10^4$, following the precursor at $t \simeq 2.2 \times 10^4$.  
After the intrusion of the two solitons, any further soliton penetration ceases for the current values of the magnetic field, and the system appears to reach a nonequilibrium steady state. 
Importantly, the solitons do not escape from the left edge during these 
processes, resulting in the increase of the total number of solitons by two. 

From the detailed simulation, we can understand the process of soliton penetration as follows. 
Due to the soliton motion to the left, the region of dominant down spins expands near the right edge, which can be regarded as the FFM state locally. 
As the energy of the FFM state is higher than the CSL in the static case, this is an unstable state and reconciled by introducing an additional soliton through the large spin precession at the edges of the system (see also the discussion on the energy barrier for a single soliton in Appendix~\ref{sec:barrier}). 
When a sufficient number of solitons are added, the system achieves a nonequilibrium steady state under the balance between the force to bring in a soliton and the repulsive forces between the neighboring solitons. 
Presented here are the results with a counterclockwise rotating magnetic field, which leads to the soliton penetration from the right edge, but employing a clockwise rotating magnetic field leads to soliton penetrations from the left edge, yet reaching a left-right inverted nonequilibrium steady state. 
This can be understood from the twofold rotation about the $z$ axis, which transforms the counterclockwise rotating magnetic field to the clockwise one. 
We note that similar soliton penetration also takes place at the boundaries of domains with different chirality. 
In this case, however, a pair of solitons is created at the boundary each of which penetrates into the different domains (see Appendix~\ref{sec:penetration_domain}).

Figure~\ref{fig:spin_xt}(b) is the result for a stronger rotating magnetic field, $h^{\rm rot}=0.012$. 
Notably, in this case, the initial penetration occurs earlier in time, the temporal interval between successive soliton penetrations gets shorter, and the total number of soliton penetrations becomes larger. 
This is understood, along with the above discussion, as follows. 
As the amplitude of the rotating magnetic field increases, a more pronounced force is exerted upon the solitons, compelling them to shift towards the left. 
This acceleration of the precursor motion contributes to the decrease in the initial set-in time of the penetration as well as the temporal interval.
Furthermore, to balance this stronger force with the repulsive force from the neighboring solitons, a greater number of solitons is required. 

\begin{figure}[tb]
\centering
\includegraphics[width=1.0\columnwidth]{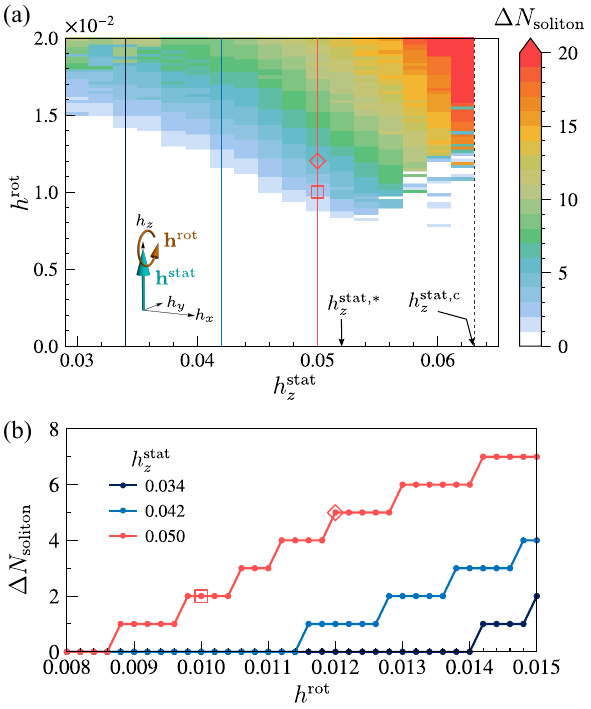}
\caption{
\label{fig:dN_hx=0}
(a) The total number of solitons penetrating into the system, $\Delta N_{\rm soliton}$ [Eq.~\eqref{eq:DeltaNsoliton}], on the plane of $\hst_z$ and $h^{\rm rot}$ with the counterclockwise rotating magnetic field at $\hst_x=0$. 
The frequency of the magnetic field is set as the edge mode in the lowest-energy magnon gap for each $\hst_z$. 
The dotted line denotes the critical magnetic field $h_z^{\rm stat,c}$, for the transition from the CSL to the FFM state. 
(b) $h^{\rm rot}$ dependence of $\Delta N_{\rm soliton}$ for several values of $\hst_z$: $\hst_z=0.034$, $0.042$, and $0.05$, corresponding to the black, blue, and red lines in (a), respectively. 
The results corresponding to Figs.~\ref{fig:spin_xt}(a) and \ref{fig:spin_xt}(b) are denoted by the red squares and diamonds, respectively. 
}
\end{figure}

While changing the values of $\hst_z$ and $\hrot$, we systematically investigate the number of solitons penetrating into the system. 
We calculate the difference of the soliton number between the initial state at $t=0$ and the final state at $t=t_f$ as 
\begin{eqnarray}
\Delta N_{\rm soliton} = \frac{1}{50T}\int_{t_f-50T}^{t_f} N_{\rm soliton}(t) dt - N_{\rm soliton}(0), 
\label{eq:DeltaNsoliton}
\end{eqnarray}
where $T=\frac{2\pi}{\omega}$ is the time period of the rotating magnetic field, and $N_{\rm soliton}(t)$ is the number of solitons given by
\begin{eqnarray}
N_{\rm soliton}(t)=\frac{1}{2\pi}\sum\limits_{l=0}^{L-2}
\mathrm{Arg}\left(\frac{S_{l+1}^{+}(t)}{S_l^{+}(t)}\right), 
\label{eq:Nsoliton}
\end{eqnarray}
with
\begin{eqnarray}
S_l^{+}(t)=S_{l,x}(t)+iS_{l,y}(t). 
\end{eqnarray}
Here, by averaging the values of $N_{\rm soliton}(t)$ during the final 50 periods of the rotating magnetic field, we eliminate the fine time dependence of $N_{\rm soliton}(t)$ and obtain the value close to an integer.

Figure~\ref{fig:dN_hx=0}(a) shows $\DNsol$ for the counterclockwise rotating magnetic field with $\epsilon=1$ on the plane of $\hst_z$ and $\hrot$. 
In this case, solitons penetrate from the right edge. By employing the clockwise rotating field with $\epsilon=-1$, solitons penetrate from the left edge, while the results of $\DNsol$ are common due to the twofold rotational symmetry about the $z$ axis for $\hst_x=0$. 
The frequency of the rotating field is set at the edge modes in the first magnon gap for each $\hst_z$. 
We find that for all $\hst_z$ the soliton penetration occurs above a certain threshold value of $\hrot$. 
Above the threshold, $\DNsol$ starts to incrementally grow in a sequential manner with $\hrot$. 
This trend is exemplified for $\hst_z=0.034$, 0.042, and 0.05 in Fig.~\ref{fig:dN_hx=0}(b). 
Such sequential increment is no longer observed for $\hrot \gtrsim 0.015$, due to strong fluctuations under large $\hrot$, as represented by the nonmonotonic color changes in Fig.~\ref{fig:dN_hx=0}(a). 
The threshold value of $\hrot$ for the soliton penetration steadily diminishes with increasing $\hst_z$, but it becomes ambiguous and seemingly increases with $\hst_z$ above $h_z^{\rm stat, *}\simeq 0.052$, where the increment of $\DNsol$ with $\hrot$ is also fluctuating.

The reduction of the threshold for $\hst_z \lesssim h_z^{\rm stat, *}$ can be attributed to the enhancement of the precession motion of spins around the edges with $\hst_z$. 
In fact, it was reported that the dynamical spin susceptibility of the edge mode monotonically increases with $\hst_z$~\cite{Shimizu2023EEF}.
Thus, larger $\hst_z$ facilitates a stronger flow of spin waves in the system, decreasing the threshold value. 
Meanwhile, the increase of the threshold for $\hst_z\gtrsim h_z^{\rm stat, *}$ can be understood from the energy experienced by a chiral soliton around the edge that is 
approximately regarded as the FFM state. 
Considering an addition of a single chiral soliton 
around the edge, for $\hst_z\lesssim h_z^{\rm stat, *}$, the energy is reduced by the addition of the soliton inside the bulk, while for $\hst_z\gtrsim h_z^{\rm stat, *}$, the system prefers to preserve the 
bulk spin configuration, limiting the modulations of spins around the edge (see also Appendix~\ref{sec:barrier} for the discussion on the energy gain accompanied by the soliton penetration into the FFM state). 
Consequently, the penetration of a single soliton 
from the edge becomes unstable for $\hst_z\gtrsim h_z^{\rm stat, *}$, whereas the significant spin precession at the edge with large $\hrot$ can force a chiral soliton to be pushed deeply inside the bulk, leading to the increase of the threshold value of $\hrot$. 

\subsection{$h_x\neq0$ case \label{sec:penetration_hx>0}}

\begin{figure}[tb]
\centering
\includegraphics[width=1.0\columnwidth]{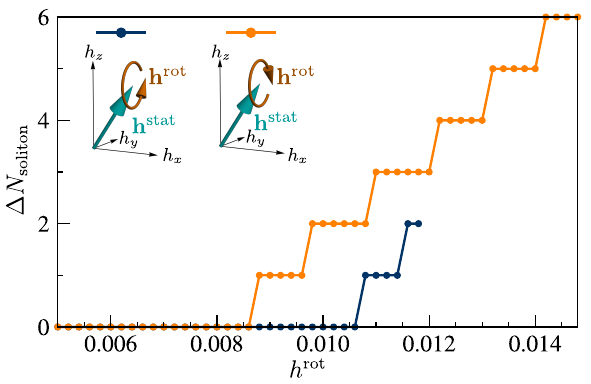}
\caption{
\label{fig:dN_hx>0}
$h^{\rm rot}$ dependence of $\Delta N_{\rm soliton}$ at $(\hst_x, \hst_z)=(0.04, 0.042)$ and $\omega=0.068$. 
The blue and orange lines represent the results for the counterclockwise ($\epsilon=1$) and clockwise ($\epsilon=-1$) rotating magnetic fields, respectively. 
The results for $h^{\rm rot}\geq0.012$ with $\epsilon=1$ are not shown as the spin texture is strongly disturbed.  
}
\end{figure}

In this section, we discuss the soliton penetration in the presence of the static magnetic field parallel to the helical axis, $\hst_x$. 
Even for nonzero $\hst_x$, the magnon band gap is opened up by $\hst_z$ and the edge modes appear within the band gap in pairs; ideally, these modes remain degenerate due to the same symmetry as in the $\hst_x=0$ case (see Sec.~\ref{sec:SW_edge}). 
However, $\hst_x$ breaks the twofold rotational symmetry about the $z$ axis, making the results by clockwise and counterclockwise rotating magnetic fields different. 
Figure~\ref{fig:dN_hx>0} shows $\DNsol$ while changing $\hrot$ at $(\hst_x, \hst_z)=(0.04, 0.042)$ for the two cases: 
The blue and orange lines represent the results for the counterclockwise ($\epsilon=1$) and clockwise ($\epsilon=-1$) fields, respectively. 
We find that the threshold values of $\hrot$ at which $\DNsol$ starts to increase become inequivalent for the clockwise and counterclockwise cases. 
Remarkably, the threshold values decrease in both cases compared to the result for $\hst_x=0$: $0.0087$ and $0.0107$ respectively for the clockwise and counterclockwise fields, are reduced from $0.0115$ in Fig.~\ref{fig:dN_hx=0}(b). 
This is due to the conical spin canting induced by $\hst_x$, which lowers the energy barrier for a chiral soliton experienced at the edge of the system. 
Upon surpassing this threshold, $\DNsol$ incrementally grows as shown in Fig.~\ref{fig:dN_hx>0}, similar to the results for $\hst_x=0$.
For the counterclockwise rotating magnetic field with $\hrot \gtrsim 0.0119$, the system experiences significant disturbance, and hence we omit the data in Fig.~\ref{fig:dN_hx>0}.


\section{Discussion \label{sec:discussion}}

In our 
monoaxial chiral magnet, the edge modes appear associated with the solitonic feature of the spin texture acquired by the static magnetic field perpendicular to the helical axis. 
Similar edge modes have also been found in a two-dimensional chiral magnet exhibiting a skyrmion lattice~\cite{Diaz2020}. 
Thus, the emergence of the magnonic edge modes is expected to be a universal feature of chiral magnets with magnetic solitons. 
Furthermore, the physical phenomena associated with these edge modes, explored in our research, are expected in chiral magnets exhibiting other magnetic solitons such as skyrmions and hopfions.

Magnetic solitons, known for demanding substantial energy to create in the bulk, render the edges of the system crucial in their creation and annihilation. 
Indeed, several mechanisms have been proposed to create solitons by leveraging boundary effects~\cite{Muller2016,Leonov2017,Ohkuma2019}. 
Meanwhile, these studies focused primarily on creating solitons within a ferromagnetic state, and the methods to introduce the desired number of solitons into a ground state, where a periodic array of solitons already exists, remained unexplored. 
The difficulty lies in repulsive forces from other existing solitons, hindering the penetration. 
In the present study, we showed that activating the edge mode with the rotating magnetic field is an efficient way to introduce additional solitons from the edge; it induces the unidirectional flow of the spin wave and the compression of chiral solitons toward the edge, leaving a soliton-sparse region around the opposite edge and enabling soliton penetration exempted from the repulsive forces. 
Such a translational motion due to spin wave propagation has been reported in various spin textures, e.g., domain walls~\cite{Han2009,Seo2011,Yan2011},  helical spin structures~\cite{Nina2021}, and skyrmions~\cite{Iwasaki2014,Schutte2014,Koide2019,Nina2023}. 
Thus, soliton infiltration is expected to occur for other magnetic solitons, similar to the existence of the edge modes. 
Our results provide valuable insights toward establishing a comprehensive theoretical framework for controlling solitons in chiral magnets.

The number of magnetic solitons is intimately linked to the magnetic and electrical properties, particularly pronounced in micro-sized samples. 
Indeed, in a monoaxial chiral magnet CrNb$_3$S$_6$ exhibiting the CSL, hystereses of the magnetization and the magnetoresistance have been observed, attributed to differences in the soliton number during magnetization and demagnetization processes~\cite{Wilson2013,Kishine2014,Togawa2015,Wang2017, Shinozaki2018,Ohkuma2019,Paterson2019}. 
The difference of the resistivity between the CSL and FFM states is roughly proportional to the number of solitons~\cite{Togawa2016,Okumura2017}. 
Our results in Fig.~\ref{fig:dN_hx=0} showed that maximally about 20 solitons can be introduced to the CSL state with 30--40 solitons, suggesting more than 50\% change of the resistivity. 
Moreover, electrons strongly coupled to such spin structures exhibit intriguing quantum transport phenomena associated with the emergent electromagnetic field brought by the Berry phase mechanism~\cite{Volovik1987,Xiao2010,Nagaosa2012-1,
Nagaosa2012-2,Nagaosa2013}. 
In chiral magnets with one-dimensional spin modulations, the unconventional inductance~\cite{Nagaosa2019,Yokouchi2020,Kitaori2021emergent} and the resonant enhancement of the emergent electric field~\cite{Shimizu2023EEF} have been discussed. 
With two-dimensional spin modulations, in addition to such emergent electric responses, the topological Hall effect arises through the emergent magnetic field originating from magnetic solitons accompanied by nontrivial topological properties~\cite{Loss1992, Ye1999, Bruno2004, Onoda2004, Binz2008}. 
Given that the magnitude of these responses is governed by the number of solitons, the dynamical method unveiled in this study for creating higher-density magnetic soliton states holds promise for enhancing the functionality of spin textures beyond that of the ground state.

Finally, let us estimate the typical timescale for soliton penetration. 
The actual timescale depends on the magnitude of the magnetic interactions specific to each material. 
Assuming a typical exchange interaction of 1 meV, the time unit $t = 1$, the frequency $\omega = 1$, and the magnetic field $h=1$ correspond to $\sim0.66$ psec, $\sim241$ GHz, and $\sim8.6$ T, respectively. 
Typically in our results, the frequency of the edge modes is $\omega \sim 0.07 \sim 17$ GHz, which is comparable to the bulk resonance frequency, and the amplitude of the rotating magnetic field required for soliton penetrations is $\hrot\sim0.01\sim860$ Oe. 
For instance, considering the cases depicted in Fig.~\ref{fig:spin_xt}, which are close to these conditions, the temporal interval between soliton penetrations is $t \sim  10^3\text{--}10^4 \sim 1\text{--}10$ nsec. 
We note that one can lower this temporal interval by increasing $\hrot$. 
The overall time for the system to reach the nonequilibrium steady state could depend on the system size. 
Assuming a typical lattice constant of 1 nm, our $L = 1000$-site system corresponds to a sample length scale of $10~\mu$m, requiring approximately $t \sim 10^4\text{--}10^5 \sim 10\text{--}100$ nsec for all solitons to penetrate. 
Therefore, the control of magnetoelectric properties associated with soliton penetration is expected to be achieved on nanosecond to microsecond timescales. 
We note that these results may also depend on the Gilbert damping, while we used a typical value for ferromagnetic materials $\alpha=0.04$~\cite{Mizukami2001, Mizukami2010,Oogane2006,Oogane2010}.


\section{Summary \label{sec:summary}}

To summarize, we have theoretically studied the spin-wave excitation in the bulk and edges of a 
monoaxial chiral magnet, and the soliton penetration phenomena caused by excitations of the edge mode. 
First, by using the linear spin-wave theory, we clarified that the magnon dispersion shows multiple gaps when the spin state acquires a solitonic feature of the CSL by the static magnetic field perpendicular to the helical axis, $\hst_z$, and that the edge mode appears within the magnon band gaps with the wave function being exponentially localized near the edge. 
We found that the localization of the edge mode gets stronger by increasing $\hst_z$, reflecting the increase of the magnon band gap. 
Furthermore, we clarified that the frequency of the edge mode in the lowest-energy magnon gap overall increases with $\hst_z$, while that in the second lowest-energy band gap decreases by increasing $\hst_z$. 
These edge modes disappear in the FFM state. 

Next, by numerically solving the LLG equation, we found the intriguing dynamics in which the solitons penetrate from the edges of the system by exciting the edge modes with a rotating magnetic field. 
In the CSL, we clarified that the additional solitons penetrate into the system from the right and left edge of the system by applying the counterclockwise and clockwise rotating magnetic field, respectively. 
We clarified that the penetration of chiral solitons takes place successively and stably through the following processes. 
First, the rotating magnetic field with the frequency of the edge mode triggers the spin precession around the edges, leading to a unidirectional spin-wave flow inside the system. 
Then, the bulk chiral solitons riding on the flow are compressed toward one edge of the system, leaving a soliton-sparse region where spins predominantly point downward around the other edge. 
Finally, an additional soliton infiltrates after the precursor phenomenon exhibiting the rapid shift of the bulk soliton at the very edge to reduce the energy in this sparse region.
We also found that the larger amplitude of the rotating magnetic field, $\hrot$, contributes to decreasing the temporal interval of penetrations and increasing the total number of infiltrated solitons $\DNsol$ in the nonequilibrium steady state.

Our findings unveil the fundamental aspects of the spin-wave excitations in both the bulk and edges, and demonstrate an efficient way of introducing additional solitons into the system by leveraging the soliton penetration dynamics associated with the edge modes in a 
monoaxial chiral magnet. 
These insights also pave the way for systematic control over magnetic and electronic functionalities, such as excitation profiles and magnetoresistance. 
In the present study, we elucidate the soliton penetration from the edges in the direction of the helical axis by using the effective one-dimensional model, while the effects of the actual spatial dimension can be important in soliton creation and annihilation. 
Indeed, it is pointed out that chiral solitons can penetrate from the sample edges in the perpendicular direction to the helical axis~\cite{Ohkuma2019}. 
The combination of soliton penetrations from different directions presumably leads to faster soliton penetration with smaller amplitudes of the rotating magnetic field. 
Furthermore, in two- and three-dimensional chiral magnets, various magnetic solitons can emerge, e.g., skyrmions 
and hopfions. 
As the edge modes can be a fundamental property in chiral magnets with magnetic solitons, it is also intriguing to extend our study to other dimensions exhibiting topological magnetic solitons.


\begin{acknowledgments}
The authors thank J. Kishine, Y. Shimamoto, Y. Togawa, and H. Watanabe for fruitful discussions.
This research was supported by Grant-in-Aid for Scientific Research Grants (Nos. JP19H05822, JP19H05825, JP21J20812, JP22K03509, JP22K13998, JP23K25816, and 24K22870), JST CREST (Nos. JP-MJCR18T2 and JP-MJCR19T3), and the Chirality Research Center in Hiroshima University and JSPS Core-to-Core Program, Advanced Research Networks. K.S. was supported by the Program for Leading Graduate Schools (MERIT-WINGS). Parts of the numerical calculations were performed in the supercomputing systems in ISSP, the University of Tokyo.
\end{acknowledgments}




\appendix

\section{Soliton penetration from chirality domain boundary 
\label{sec:penetration_domain}}

\begin{figure}[h]
\centering
\includegraphics[width=1.0\columnwidth]{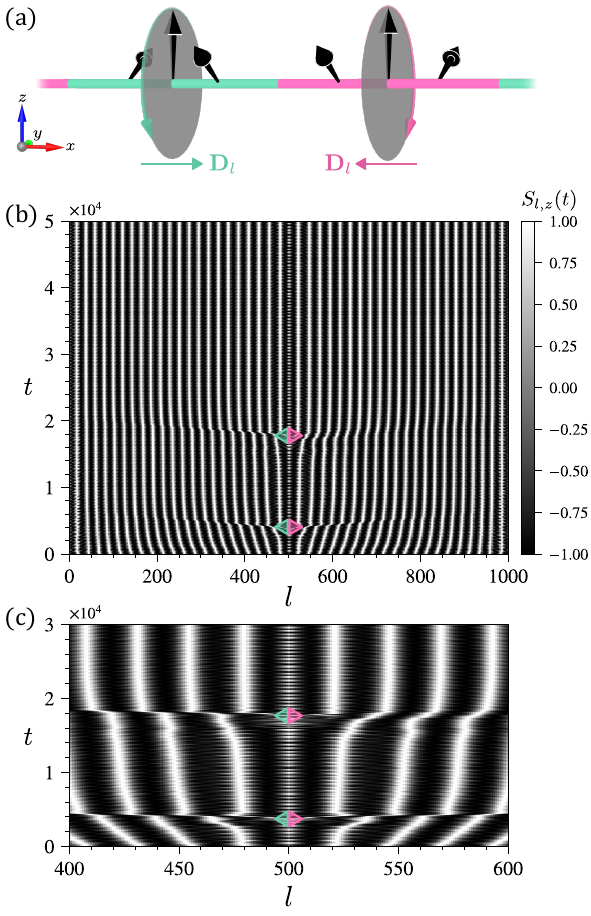}
\caption{
\label{fig:penetration_domain}
(a) Schematic picture of the setup for the system with the domain of the opposite chirality in Eq.~\eqref{eq:ham_domain} with Eq.~\eqref{eq:DM_domain}. 
The black arrows represent the spins with the opposite swirling directions between the green and pink regions. 
(b) Real-space and real-time distribution of the $z$ component of spins $S_{l,z}(t)$ for the 1000-site system with the counterclockwise rotating magnetic field for $(\hst_x, \hst_z)=(0, 0.05)$, $\omega=0.076$, and $h^{\rm rot}=0.013$. 
(c) The enlarged plot for 200 sites at the center of the system with a shorter time window. 
The green and pink triangles represent the time when solitons penetrate simultaneously into the left and right domains of the system, respectively. 
}
\end{figure}

In this Appendix, we demonstrate that soliton penetration dynamics at the edges of the system 
discussed in Sec.~\ref{sec:penetration} takes place at a boundary between domains with different chirality. 
Here, we consider the system where crystal structures with distinct chiralities form domains. 
Such systems have indeed been observed in experiments, wherein micro-sized samples exhibit the spin texture coexisting with right-handed and left-handed CSLs due to the opposite chirality, by using Lorentz transmission electron microscopy~\cite{Togawa2015}. 
To mimic such situations, we consider a $1000$-site system under the PBC, where the sign of the DM interaction is reversed in the right half of the system such that the left half exhibits a right-handed CSL, while the right half displays a left-handed CSL, as schematically depicted in Fig.~\ref{fig:penetration_domain}(a). 
The Hamiltonian reads 
\begin{align}
\mathcal{H}(t)=
\sum_l \bigl[ &-J{\bf S}_l(t)\cdot{\bf S}_{l+1}(t) 
-{\bf D}_l \cdot \left( {\bf S}_l(t) \times {\bf S}_{l+1}(t) \right) \notag \\ 
&+ {\bf h}(t)\cdot{\bf S}_l(t) \bigr], 
\label{eq:ham_domain}
\end{align}
where $D=|{\bf D}_l|=\tan\left(\frac{\pi}{10}\right)$ and  
\begin{align}
{\bf D}_l=
\begin{cases}
D\hatv{x} & (0 \leq l < 500),\\
-D\hatv{x} & (500 \leq l < 1000).
\end{cases} 
\label{eq:DM_domain}
\end{align}
We calculate the spin dynamics by numerically solving the LLG equation in Eq.~\eqref{eq:LLG} with the same parameters used in Sec.~\ref{sec:penetration}.

This system exhibits the mid-gap modes with the same frequency as the edge mode in Sec.~\ref{sec:SW_edge}, and these modes are exponentially localized at the boundaries of the different chirality, i.e., $l\simeq0$ and $l\simeq500$. 
In Fig.~\ref{fig:penetration_domain}(b), we show the real-space and real-time distribution of $S_{l,z}(t)$ with $(\hst_x,\hst_z)=(0, 0.05)$ for the counterclockwise rotating magnetic field with $\omega=0.076$ and $h^{\rm rot}=0.013$. 
After applying the rotating magnetic field, the system exhibits the precursor motion wherein the bulk solitons in the left (right) half are compressed toward $l=0$ ($l=L-1$), and accordingly, the soliton-sparse region is realized around $l=500$. 
Subsequently, additional solitons penetrate into the system from the boundary of the different chirality at $t\simeq4000$, as indicated by the green and pink triangles. 
Notably, solitons do not penetrate into just one chirality domain but both domains simultaneously, as seen in the enlarged figure in Fig.~\ref{fig:penetration_domain}(c).   
This is linked to spatial inversion symmetry of the system at the boundary. 
We also find another soliton penetration at $t\simeq1.8\times10^4$ in a similar manner.

\section{Surface barrier \label{sec:barrier}}

\begin{figure}[tb]
\centering
\includegraphics[width=1.0\columnwidth]{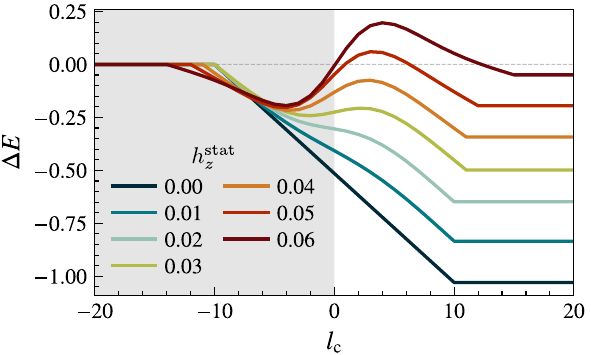}
\caption{
\label{fig:surface_barrier}
Energy for a single chiral soliton measured from the fully-polarized FFM state, $\Delta E$, while changing the center position of the soliton $l_{\rm c}$ with $\hst_x=0$ for several values of $\hst_z$. 
The white and gray regions represent the inside and outside of the $1000$-site system, respectively. 
}
\end{figure}

\begin{figure}[tb]
\centering
\includegraphics[width=1.0\columnwidth]{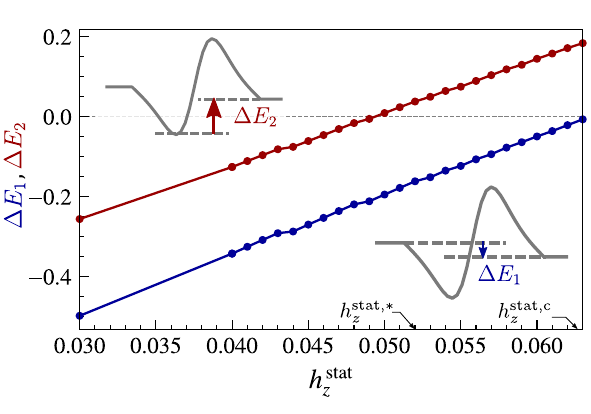}
\caption{
\label{fig:barrier_hzdep}
The energy gain $\Delta E_{\rm 1}$ achieved by introducing a single soliton into the FFM state, and the optimal energy difference $\Delta E_{\rm 2}$ when placing the soliton either outside or inside the FFM state, while varying $\hst_z$, represented by the blue and red lines, respectively. 
See Eqs.~\eqref{eq:app_DE1} and \eqref{eq:app_DE2} for the definitions of $\Delta E_1$ and $\Delta E_2$, respectively. 
}
\end{figure}

In this Appendix, we investigate a potential barrier encountered when introducing a single chiral soliton from an edge of the system into the bulk of the FFM state. A similar situation was examined in a continuum model~\cite{Shinozaki2018}. 
We first prepare a single chiral soliton capped by polarized spins on both sides. 
Specifically, we prepare the system of size $L+1$ with ${\bf S}_{0}={\bf S}_{L}=(0,0,-1)^{\mathsf T}$ and optimize the spin texture based on the Hamiltonian in Eq.~\eqref{eq:ham}.
The optimization is conducted while varying $L$ from 20 to 40 one by one, and we identify the lowest-energy single chiral soliton state. 
Subsequently, we embed this single chiral soliton state to be centered at $l = l_{\rm c}$ into the fully-polarized FFM state with ${\bf S}_l=(0,0,-1)^{\mathsf{T}}$ for the $L=1000$-site system ($l=0,1,\ldots,L-1$) and calculate the total energy $E_{l_{\rm c}}$ while varying $l_{\rm c}$ from $-20$ to $20$. 
If a part of the soliton extends beyond the edge of the system when $l_{\rm c}$ approaches zero or becomes negative, we compute the energy by disregarding the protruding portion. 
These analyses facilitate calculating the potential barrier encountered for the single soliton to penetrate from the edge into the FFM system. 

Figure~\ref{fig:surface_barrier} shows the $l_{\rm c}$ dependence of the total energy measured from the fully-polarized FFM state, namely, $\Delta E = E_{l_{\rm c}} - E_{l_{\rm c}=-20}$, for several $\hst_z$ with $\hst_x=0$. 
For $\hst_z=0$, $\Delta E$ decreases by increasing $l_{\rm c}$, meaning that the chiral soliton can smoothly penetrate into the system without any potential barrier, and finally, the system gains the energy of $\Delta E \simeq -1.03$ for sufficiently large $l_{\rm c}$. 
By increasing $\hst_z$, the amount of this energy gain decreases, and the slope of $\Delta E$ at $l_{\rm c}=0$ increases. 
The slope becomes zero at $\hst_z\simeq 0.025$ and then turns to be positive by further increasing $\hst_z$, resulting in a dip and a hump for $l_{\rm c}<0$ and $l_{\rm c}>0$, respectively. 
The hump corresponds to the potential barrier experienced by the single chiral soliton. 
Although the system still gains the energy by introducing the chiral soliton, i.e., $\Delta E_{l_{\rm c}=20}$ is negative, the soliton is required to overcome the potential barrier around the edge.  
The energy gain achieved by the soliton penetration turns to be positive for $\hst_z \gtrsim h_z^{\rm stat, c}$ as the FFM state is the ground state. 
Our results for the discrete lattice system are qualitatively consistent with the previous study in continuum space~\cite{Shinozaki2018}.

In Fig.~\ref{fig:barrier_hzdep}, we show the $\hst_z$ dependence of 
the energy gain accompanied by introducing a soliton into the fully-polarized FFM state, $\Delta E_1$, defined by
\begin{eqnarray}
\Delta E_1=E_{l_{\rm c}=20}-E_{l_{\rm c}=-20}, \label{eq:app_DE1}
\end{eqnarray}
and the optimal energy difference between the cases where the soliton is placed outside ($l_{\rm c}\leq0$) and inside ($l_{\rm c}\geq0$) the system, $\Delta E_2$, defined by
\begin{eqnarray}
\Delta E_2=\underset{l_{\rm c}\geq0}{\min}\Delta E_{l_{\rm c}}
-\underset{l_{\rm c}\leq0}{\min}\Delta E_{l_{\rm c}}.  
\label{eq:app_DE2}
\end{eqnarray}
See the insets for the definitions. 
As discussed above, the energy gain $\Delta E_1$ is negative at $\hst_z=0$,  monotonically increases with $\hst_z$, and becomes zero at $h_z^{\rm stat, c}\simeq0.063$. 
Similarly, $\Delta E_2$ also starts from negative and increases with $\hst_z$, whereas it turns to be positive for $\hst_z \gtrsim 
0.0495$. 
For $\hst_z \lesssim 
0.0495$, placing a chiral soliton inside the bulk FFM state rather than outside decreases the energy. 
Meanwhile, for $\hst_z \gtrsim 
0.0495$, it is more favorable to place a chiral soliton outside the system ($l_{\rm c} \simeq -3$). 
This sign change of $\Delta E_2$ can explain the unstable soliton penetration in Sec.~\ref{sec:penetration}.  
By applying a rotating magnetic field, the bulk solitons are compressed toward the edge of the system, leaving the sparse region at the opposite edge.  
In the soliton sparse region, spins predominantly point downward, which can be approximately regarded as the FFM state, and thus the systematic soliton penetration is expected to be hindered for $\hst_z \gtrsim 0.0495$. 
Indeed, in Fig.~\ref{fig:dN_hx=0}(a), the penetration of solitons becomes more difficult in the similar magnetic field range $\hst_z \gtrsim h_z^{\mathrm{stat,*}}\simeq0.052$. 
Meanwhile, the application of the rotating magnetic field with a larger amplitude may cause a soliton penetration due to the large disturbances in the nonequilibrium state. 
These explain the nonmonotonic behavior of the threshold value of $h^{\rm rot}$ with $\hst_z$ in Fig.~\ref{fig:dN_hx=0}(a).

The above discussions have been limited to the situation where placing a single chiral soliton in the fully-polarized FFM state. 
Nevertheless, considering the problem with multiple solitons, placing them inside the system still yields a larger energy gain than placing a single soliton outside. 
By introducing $n$ solitons deeply inside the system, the system gains the energy of $n\Delta E_1$. 
For instance, at $\hst_z=0.06$, $n\Delta E_1 \leq \Delta E_2$ is realized for $n\geq3$. 
Hence, placing multiple chiral solitons inside the bulk rather than placing a single chiral soliton outside decreases the energy for $\hst_z > h_z^{{\rm stat},*}$. 
To cause such multiple penetrations, a substantial rotating magnetic field is required, leading to the larger threshold field of $\hrot$, as shown in Fig.~\ref{fig:dN_hx=0}(a).


\bibliography{ref}

\end{document}